# Multi-Pulse Terahertz Spectroscopy Unveils Hot Polaron Photoconductivity Dynamics in Metal-Halide Perovskites

*Xijia Zheng,[†§] Thomas R. Hopper,[†§]\* Andrei Gorodetsky,[†§] Marios Maimaris,[†] Weidong Xu,[†] Bradley A. A. Martin,[‡] Jarvist M. Frost[‡] and Artem A. Bakulin[†]*

† Department of Chemistry and Centre for Processable Electronics, Imperial College London, London W12 0BZ, United Kingdom

‡ Department of Physics, Imperial College London, Exhibition Road, London SW7 2AZ, United Kingdom

\*Corresponding author e-mail: t.hopper16@imperial.ac.uk

**ABSTRACT**

Hot carriers in metal-halide perovskites (MHPs) present a foundation for understanding carrier-phonon coupling in the materials as well as the prospective development of high-performance hot carrier photovoltaics. Whilst the carrier population dynamics during cooling have been scrutinized, the evolution of the hot carrier properties, namely the mobility, remain largely unexplored. Here we introduce novel ultrafast visible pump – infrared push – terahertz probe spectroscopy to monitor the real-time conductivity dynamics of cooling carriers in methylammonium lead iodide. We find a mobility decrease upon optically re-exciting the carriers, as expected for band-transport. Surprisingly, the conductivity recovery is incommensurate with the hot carrier population dynamics measured by infrared probe and exhibits a negligible dependence on the hot carrier density. Our results reveal the importance of highly-localized lattice heating on the hot carrier mobility. This collective polaron-lattice phenomenon may contribute to the unusual photophysics of MHPs and should be accounted for in hot carrier devices.

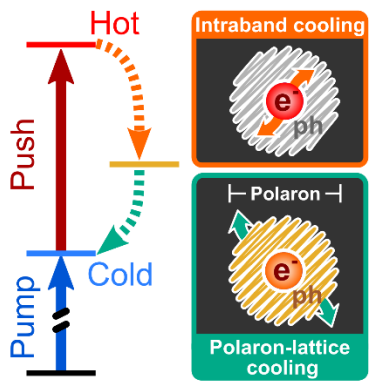

Metal-halide perovskites (MHPs) are the subject of intense optoelectronic research on account of their efficient light-absorbing and emitting characteristics,[1,2] composition-tunable bandgaps,[3] relatively high carrier mobilities[4,5] and surprisingly long carrier lifetimes/diffusion lengths.[6–8] These outstanding optical and electronic properties have prompted exhaustive investigations into the fundamental nature of the electronic excitations.[9,10] A common finding across these works is the importance of strong carrier-phonon coupling, which can be understood as being due to the soft and polar nature of the MHP lattice.[11] By and large, the onset of carrier-phonon coupling in semiconductors occurs on the ultrafast timescale where the electronic and structural degrees of freedom are far from equilibrium.[12,13] The nature of this coupling is therefore particularly salient to carrier cooling and polaron formation; two processes wherein electronic excitations exchange their energy with vibrations of the lattice to minimize the overall energy of the system.

Research into these interrelated phenomena in MHPs has been extremely active over the past decade. The remarkable properties of highly energetic (hot) carriers in MHPs were first identified in the photophysical studies that followed the discovery of highly-efficient MHP solar cells.[14,15] One of these early characterizations by Xing et al. highlighted that the lifetimes of hot carriers in MHPs spanned hundreds of femtoseconds, well beyond the timescale of internal vibrational relaxation in most organic semiconductors.[7] Even longer hot carrier lifetimes have been reported for MHPs, which has been ascribed to the screening of the energy losses via carrier-phonon scattering (cooling) by the polar response of the lattice surrounding the carriers (polarons).[16,17] A great deal of reports have shown that cooling can be further protracted at high excitation densities, (re)invigorating discussions about the potential harvesting of hot carriers to surpass the Shockley-Queisser efficiency limit of single-junction photovoltaics.[18–20] The slowed cooling effect at high excitation density is usually explained through some manifestation of a "hot-phonon bottleneck", where the cooling of many hot carriers produces an excess of non-equilibrium phonons that are continually reabsorbed by the charges.[21–23] A growing number of works implicate polarons in this aspect of the hot carrier behavior too, owing to the enhanced spatial extent, strong coupling to phonons and sharing of phonon subpopulations inherent to overlapping polarons.[24–27]

The interpretation of the transient optical signals that convey the aforementioned relaxation processes in MHP-based materials must be taken with care, especially when pumping far above the bandgap and at high fluences. These conditions can favor a myriad of processes that obfuscate the intraband relaxation, including, but by no means limited to: interband transitions between distinct electronic/spin states,[28–30] exciton formation/dissociation,[13,31–33] energy/charge migration,[34–36] bandgap renormalization,[21,37] refractive index changes,[21,38] Stark shifting,[39,40] and many-body (Auger) processes.[41] The interpretation is complicated further still by the lack of established analysis procedures for even the most widely used transient absorption (pump-probe, PP) methods.[42,43] To circumvent these challenges and provide new insight into the carrier relaxation dynamics of MHPs we recently developed a visible pump – infrared push – infrared probe experiment (PPP-IR) to pump the material into the lowest-energy (cold) excited state, excite ("push") these carriers into a higher-energy (hot) state optically and probe their intraband relaxation.[25] This experiment allowed us to establish the material parameters that control cooling in MHP-based materials, and also demonstrate the interplay between carrier-carrier and carrier-phonon relaxation mechanisms.[44,45] The technique has since been adapted to elucidate the extent and dynamics of hot carrier trapping[46] and extraction[47] in MHPs.

While all-optical PP and PPP approaches have been successful in unravelling the population dynamics and density of state distribution for hot carriers in MHPs, they do not provide explicit information on the properties of the carriers and lattice during relaxation. This could potentially be realized by monitoring the mobility of the carriers and the structure of the lattice as function of time. Recent developments in ultrafast microscopy[34–36] and X-ray/electron diffraction approaches[48,49] are highly encouraging in this regard, but tend to require considerable investment both in terms of experimental resources and data analysis.

A robust table-top method to ascertain the evolution of carrier mobilities and associated lattice response in semiconductors is time-resolved THz spectroscopy. This is another form of transient absorption spectroscopy employing ultrashort (sub-ps) pulses, but with a low energy probe (on the order of meV). The electric field of this THz radiation accelerates the motion of (unbound) charge carriers, thus providing a means to track the electrical mobility of the photoexcited states, as well as resolving the low-energy phonons in the material that can couple to the carriers.[50,51] The technique is commonly employed to study charge recombination and transport in MHPs.[4,52–55] Early reports focusing on the time-domain data established moderate mobilities (on the order of 10 $cm^2$ $V^{-1}$ $s^{-1}$) and slow bimolecular recombination constants for the materials.[4] Ostensibly, these aspects signify the existence of polarons, but contradictory values for the mobilities have surfaced depending on the conditions of the experiment and sample,[56] and effects such as photon recycling have also been shown to play an important role in the observed dynamics.[54] More recent studies have focused on the sub-ps timescale with a view to explore the carrier cooling dynamics.[57–60] Generally, these reports echo the results of the aforementioned transient absorption studies in that they show a delayed build-up of the signal for higher energy excitations. Bretschnedier et al. showed that this build-up cannot be solely attributed to carrier cooling, and contains an intrinsic term for polaron formation (~400 fs in the case of $MAPbI_3$).[58] On the other hand, frequency-resolved THz measurements have pointed toward the emergence of polarons in the form of a non-Drude spectral response[55,61] and/or coherent beating at particular resonances,[62–64] but are seldom able to fully distinguish this self-localization of the charge from the cooling of the carriers.

Herein, we introduce a novel ultrafast visible pump – infrared push – THz probe pulse sequence (PPP-THz) to combine the advantages of pump-push-probe and THz spectroscopies. We observe a decrease of carrier mobility upon optically energizing the carriers. This agrees with previous THz studies, and the general theoretical understanding that these materials exhibit band transport. Intriguingly, the reduction in carrier mobility does not happen instantly after the IR push, and the subsequent recovery takes longer overall (~800 fs) than the observed intraband carrier relaxation measured by PPP-IR (a few hundred fs). The THz conductivity also shows a negligible push fluence dependence, whereas the strong push fluence dependence in PPP-IR has been ascribed to the hot-phonon bottleneck effect.[25] We interpret this behavior as a two-step relaxation which includes intraband carrier cooling leading to a local (polaron) lattice heating effect. This local lattice heating is short-lived but has a substantial effect on the mobility of cooling carriers.

In this work we chose to study the prototypical MHP methylammonium lead iodide (MAPbI$_3$) as a thin film on a THz-transparent $z$-cut quartz substrate (see Supporting Information for full sample preparation details). Although better-performing MHP compositions now exist for optoelectronic applications,[2] MAPbI$_3$ is still used as a workhorse for spectroscopic investigations. Figure S1 shows the absorption and photoluminescence spectra for the sample. Based on the overlapping absorption onset and emission peak of the sample, we estimate the optical bandgap to be around 1.6 eV, in accordance with literature values for the material.[3]

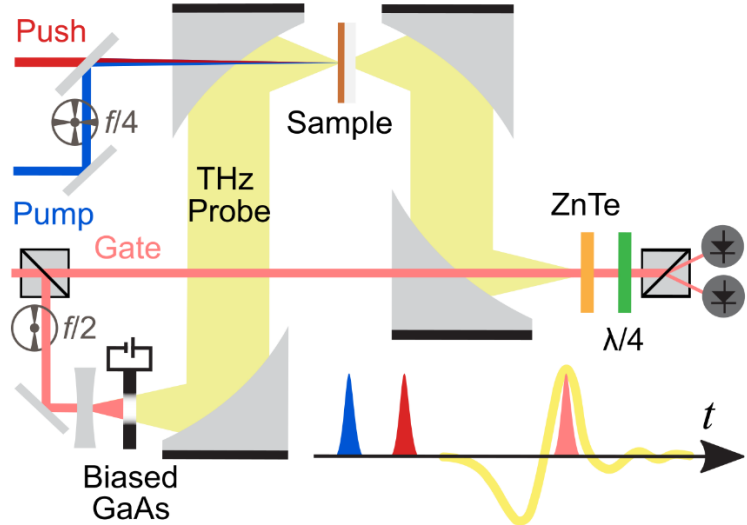

**Fig. 1.** *Simplified layout for the visible pump – infrared push – THz probe experiment. The pump and gate beams are modulated at f/4 and f/2 respectively, where f is the laser repetition frequency. The pump-push-probe pulse sequence is depicted at the bottom right.*

To explore the photoexcitation dynamics in the MAPbI₃ sample we complement the PPP-IR experiment presented in our previous work[25] with a new variant of time-resolved THz spectroscopy, PPP-THz. Figure 1 depicts the latter experimental setup. The full technical details are elaborated in the Supporting Information, but in essence there are four beams of femtosecond pulses to consider. The first three are typical of time-resolved THz spectroscopy rigs based on commercial femtosecond regenerative amplifiers, and comprise of: (i) a visible pump pulse, used to photoexcite the sample; (ii) a THz probe pulse, used to interrogate the behavior (i.e. population and conductivity dynamics) of the photoexcitations; and (iii) the gate pulse, used to modulate the electro-optic detection. Each point in the time domain of the THz waveform contributes to every frequency of the THz radiation in Fourier space. By fixing the gate delay to the peak of the THz waveform, as shown in Figure 1, one can obtain the weighted-average or frequency-integrated response as a function of the pump-probe time delay. Alternatively, the gate can be scanned along each time point in the THz waveform to relay the full transient spectral response as a function of the pump-probe delay.[50] For simplicity, we will focus on the former frequency-integrated scheme, and return to the latter frequency-resolved results toward the end of this report. The ultrashort and broadband nature of the THz probe pulses are corroborated by Figures S2 and S3, which evince the ~130 fs instrument response function (IRF) of the setup and ~15 THz bandwidth of the THz radiation, respectively.

Figure S4 shows the frequency-integrated pump-probe kinetics as a function of the pump fluence. Unless specified otherwise, all experiments in this work were performed at low pump fluences (<60 µJ cm$^{-2}$) where the early (ps) and late-time (ns) amplitudes match and scale linearly with the pump fluence. For higher intensity excitation (corresponding to carrier densities > $3 \times 10^{18}$ cm$^{-3}$) we observe prominent signatures of Auger recombination in the time-resolved data, and find that the overall dynamics are dominated by many-body interactions between carriers. To this end, we have previously experimentally established that the collisions between hot and cold carriers can in fact accelerate the intraband relaxation dynamics in MHP materials.[44,45] Our ability to restrict the cold carrier population to ignore the contributions from these processes is not practicable for standard transient spectroscopies deploying a single excitation pulse.

The unique feature of our experiment is the IR push beam spectrally tuned to a broad photoinduced absorption by the electrons and holes in MAPbI$_3$.[65,66] This beam arrives at the sample after the pump to optically re-excite the relaxed "cold" states with appreciable excess energy (600 meV) without populating higher-lying electronic bands,[28,37,67] thereby providing a means to control the number of hot states and directly track the intraband relaxation dynamics. Figure 2(a) shows that the introduction of the push ~12 ps after the pump arrival momentarily reduces absorption of the THz probe, which subsequently recovers within 1 ps. As an additional control measurement we varied the delay between the pump and push pulses but observed no effect on the recovery dynamics (see Figure S5). This indicates that the moderate excess energy (~300 meV) from the above-gap pump (1.9 eV) is dissipated before the push arrives.

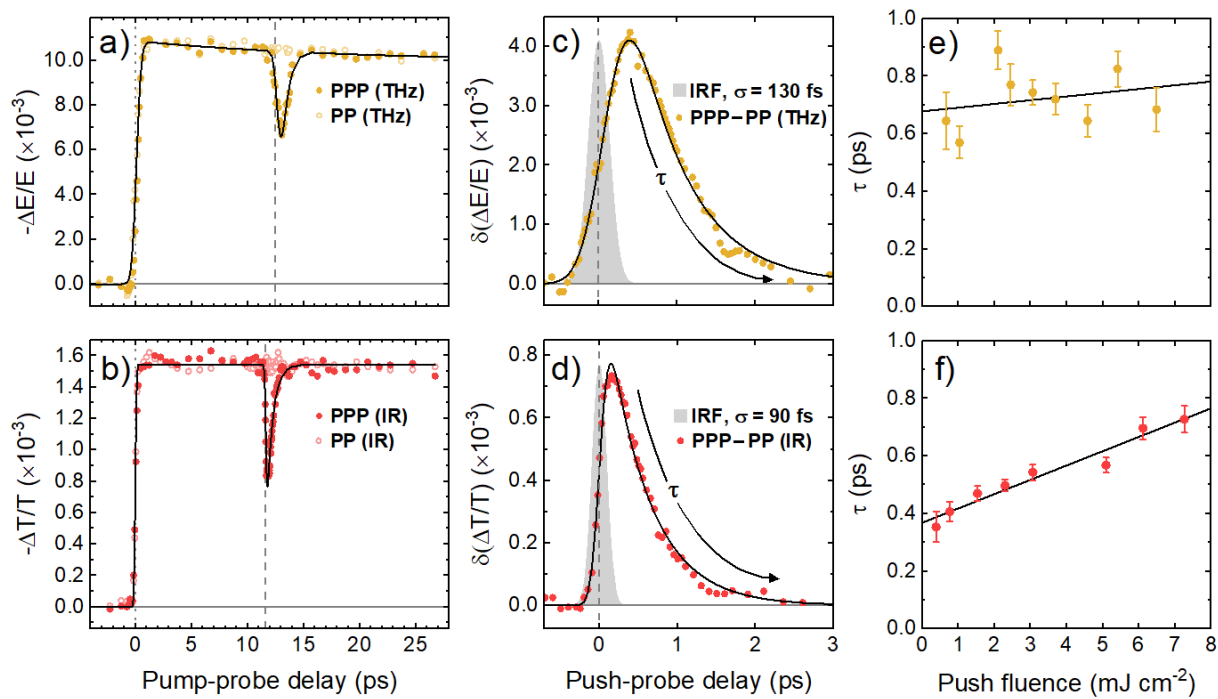

**Fig. 2.** *Representative pump-push-probe transients (left), push-induced kinetics (middle) and overall probe recovery dynamics (right) for the PPP experiments with THz and IR probes (top and bottom, respectively). The grey-shaded areas depict the Gaussian instrument response function for the experiments.*

The overall action of the push pulse on the THz absorption is qualitatively similar to what we observe in the analogous PPP-IR experiments[25,44,45] where the probe is resonant with the push transition (i.e. IR push and probe pulses have the same energy). This is shown in Figure 2(b) for the sample under study. To scrutinize the differences between the THz and IR data, we compare the push-induced kinetics by subtracting the pump-probe traces (PP) for each experiment from the corresponding pump-push-probe transients (PPP). There are two key differences between the THz and IR responses in Figures 2(c-d). Firstly, the IR response emerges within the IRF, whereas the THz response lags behind by at least 100 fs. Secondly (for low push fluences), the overall probe recovery time (τ) to the pre-push condition is almost twice as long in the THz case. Figures 2(e-f) reveal a third discrepancy in the responses when τ is measured as a function of the push fluence; the THz dynamics barely change, whereas the IR dynamics do.

From our prior works on MHPs using the PPP-NIR experiment, we can assign the absorption of the IR probe as the intraband absorption by cold carriers. We can also interpret the prompt (within IRF) change in the signal due to the push as the impulsive (re)-excitation of the cold carriers into a higher-energy hot state. The amplitude of the push-induced signal increases with the incident push fluence, and for the highest push fluence, the NIR probe absorption is almost totally depleted by the push (Figure S6), attesting to the depopulation of the cold state. This cold state and the IR absorption are restored as the carriers cool. The slowing of the carrier cooling dynamics at higher push fluences (hot carrier densities) can be understood through the framework of the hot-phonon bottleneck mechanism,[21–23] where the energy deposited into phonons through cooling is continually re-absorbed by overlapping polarons.[24–27]

Generally, the low-energy THz probe ought to be sensitive to all carriers (cold and hot), regardless of their energy within the electronic band. The push is not capable of interband excitation and therefore does not change the overall concentration of the photoexcited carriers. Since the absorption of THz radiation by charges is sensitive to both their concentration and mobility, it follows that the reduced THz opacity caused by the push is due to the lower mobility of the pushed (hot) carriers compared to the unpushed (cold) carriers. The reduced mobility of the hot states may be due to an increase in their effective mass, as outlined in recent works.[60,68] However, taking the delayed response of the THz dynamics following the IR push, we suggest that this reduced mobility is instead caused by a temporary localized heating[69] of the MHP lattice, associated with the immediate vicinity of the localized charge-carrier (polaron). The small volume of this polaron leads to a high effective lattice temperature, and substantial reduction of the carrier mobility. On a timescale of ps, heat flows out of the polaron volume (equivalently, the polaron moves) and the effective temperature equilibrates with the surrounding ambient lattice, thus restoring the mobility to the pre-push level. This thermal process causes the lifetime of the signal recovery to be constant regardless of the push fluence. Unlike in the NIR experiment, the THz probe absorption is never fully "bleached" by the push, and instead gives an overt saturation (Figure S6). From this, the push induced reduction in mobility can be quantified from the THz response as being a 35% reduction. From our previous model of temperature-dependent mobility,[68] this reduction from ambient occurs at an effective temperature of 700 K. Equating the 0.6 eV push with this 400 K uplift, we can estimate that there are 18 degrees of freedom which make up the local heat bath around the charge carrier.

To capture the dynamics from the experimental data in a more quantitative manner, we propose the three-state kinetic model depicted in Figure 3(a). Here, cold polarons ($n_c$) are re-excited by the push to form hot polarons ($n_h$). The hot polarons undergo intraband cooling, transferring the excess energy from the carrier to phonons within the polaron volume to form the hot polaron lattice state ($n_{hl}$). The system returns to the initial cold polaron state after the local lattice heat is equilibrated with the lattice beyond the polaron volume. We emphasize that, though this kinetic model is ostensibly similar to one proposed in the work of Bretschneider et al.,[58] the nature of intermediate state is different. Their model concerns the relaxation of the pre-polaron (free carrier) state and synchronous formation of the polaron, which has also been proposed in other notable works.[16,70] In our experiment, we assume that initial pump pulse forms the polaron several ps before it encounters the push. In the case that polaron formation in MAPbI₃ takes ~400 fs, it is unlikely to be 'unformed' within a ~100 fs push pulse based on the principle of microscopic reversibility. We therefore exclude a polaron formation time in our analysis and attribute the observed relaxation processes to intraband cooling and polaron-lattice cooling. The system of rate equations for the three-state model is shown below. We assume that PPP-IR tracks the $n_h$ population while PPP-THz principally follows the $n_{hl}$ population.

$$\frac{dn_h}{dt} = I_{push}(t) \cdot n_c - k_c \cdot n_h \tag{1}$$

$$\frac{dn_{hl}}{dt} = k_c \cdot n_h - k_l \cdot n_{hl} \tag{2}$$

$$\frac{dn_c}{dt} = -I_{push}(t) \cdot n_c + k_l \cdot n_{hl} \tag{3}$$

Figure 3(b) shows representative global modelling of PPP-IR and PPP-THz data obtained at the same pump and push fluence. The full datasets obtained across a range of push fluences are displayed in Figure S7. In all cases, the delayed rise of the PPP-THz signal is well reproduced by the intraband cooling process that is observed as a decay of PPP-IR. The summary of extracted time constants for all push fluences is presented in Figure 3(c). Interestingly, the model suggests that polaron-lattice cooling has a constant timescale around 450 fs regardless of the $n_h$ population density. Meanwhile, the carrier intraband cooling timescale $k_c$ increases with this population density from 400 to 650 fs, as was previously observed and attributed to the hot-phonon bottleneck effect.[21–23,25–27,43] The observed timescales of both cooling processes are in excellent agreement with Bretschneider et al.[58] However, the fact that a ~450 fs local lattice cooling timescale $k_l$ is observed in our experiment where polaron formation precedes (and is therefore ruled out from) the observed dynamics may point toward an alternative interpretation of their results.

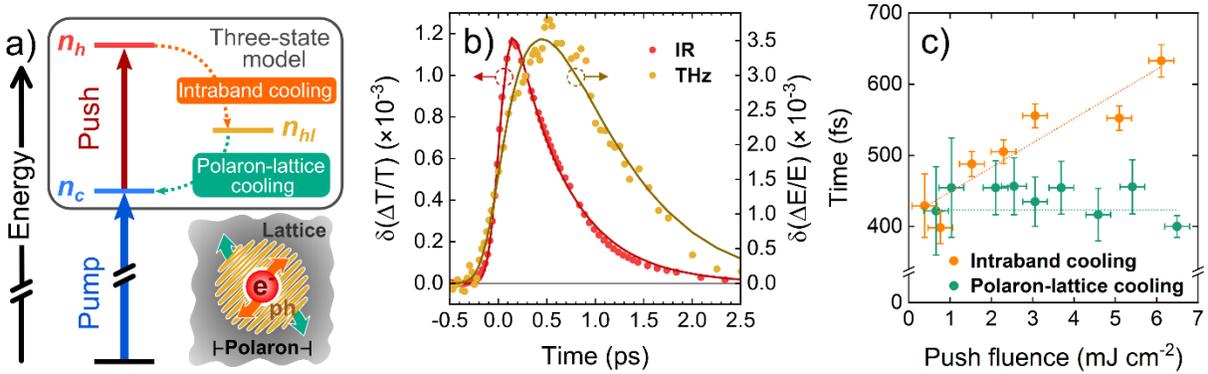

**Fig. 3.** *(a) Energy diagram of the three-state model used to interpret the PPP-THz data. $n_c$, $n_h$ and $n_{hl}$ represent the cold polaron, hot polaron and polaron-hot lattice populations, respectively. The inset provides a microscopic depiction of the model, where $e^-$ and ph represent an electron and the surrounding phonons. (b) Experimental PPP-IR and PPP-THz push-induced transients at comparable push fluences (5.1 and 5.4 mJ cm$^{-2}$, respectively), with fits from the system of coupled rate equations. (c) Intraband cooling and polaron-lattice cooling times extracted from the model fits of the experimental data (see Figure S7). Dotted lines are linear fits as a guide to the eye.*

To verify our model further, we perform the frequency resolved PPP-THz experiments (scanning both the THz probe and gate delays at each pump delay) to reconstruct the evolution of the dielectric response during carrier relaxation. The results are shown in Figure 4(a). Upon pumping the sample we observe two strong resonances at ~1.25 and 2.25 THz which can be attributed to two groups of optical phonons associated with the bending and stretching of the Pb-I bonds that strongly couple to the carriers.[52,53,55,71–74] When the push arrives (in this case ~10 ps after the pump), the THz absorption across the whole spectrum is diminished, with the most pronounced reduction at the two aforementioned peaks. This is also shown in Figure 4(b), which breaks down the complex photoconductivity spectrum into real and imaginary components. The non-Drude like spectral response tentatively precludes free (non-polaronic) carriers as being the sole origin of the signal.[55,61] The evolution of this response as a function of time is non-trivial and will likely require an elaborate analysis of the relevant carrier-phonon couplings at various degrees of quasi-thermal equilibration which is beyond the scope of this work. Notwithstanding the rich spectral behavior, the dominant effect is a drop in conductivity upon pushing the electronic states to a higher energy level. The interplay between the polaron dynamics and lattice is further corroborated by Figure 4(c), which shows that the conductivity recovery dynamics are strongly dependent on the THz probe frequency, with the slowest dynamics patently occurring at the peaks. This observation is fairly intuitive in the case that the underlying phonons are strongly coupled to the carriers. In such a scenario, the phonons will be directly heated by the carriers during cooling and before the heat can subsequently dissipate from the modes by scattering into other (non-coupled) phonons. This picture is consistent with our model which essentially conveys the physics as a system of coupled heat baths.

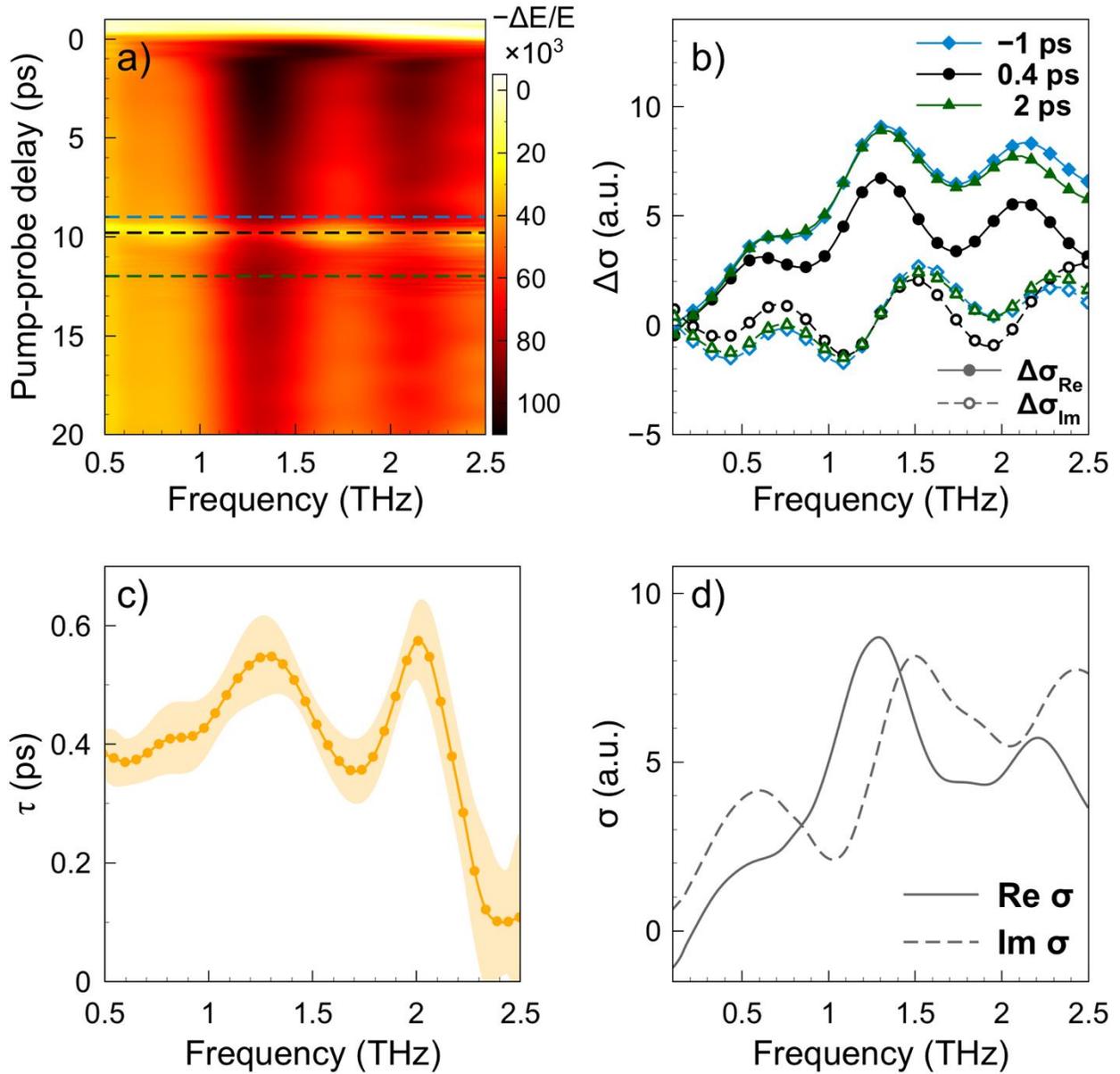

**Fig. 4.** *(a) 2D probe frequency/push-probe delay intensity map for the visible pump-IR push-THz probe experiment, with the pump-push delay set to 10 ps. (b) THz photoconductivity spectra showing the real and imaginary components of the complex conductivity ($\Delta\tilde{\sigma} = \Delta\sigma_{Re} + \Delta\sigma_{Im}$) before, at and after the arrival of the push (as depicted by the blue, black and green dashed lines on the intensity map to the left). (c) Overall probe recovery time following the push at different probe frequencies. (d) Simulated frequency-resolved conductivity from the FHIP polaron-mobility theory[75] and an extended version of the material-specific model.[68] The real-component maxima at the phonon group frequencies of 1.25 and 2.25 THz compare well with the measured response, and correspond to a polaron-renormalized interaction with the underlying carrier-phonon coupled modes.*

By calculating the full frequency-dependent electrical susceptibility of the polaron in the Feynman-Hellwarth-Iddings-Platzman (FHIP) model,[75] we can simulate the THz response of our lead-halide perovskite model.[68] This leads to reasonable qualitative agreement with the observed response. Interestingly we found the best agreement with an effectively zero temperature mobility theory, suggesting that the sub-ps timescale of the THz measurement corresponds to a pre-thermal equilibrium mobility regime. An extended analysis of the frequency-dependent response will be the subject of a future investigation.

In summary, we have developed a novel visible pump – infrared push – terahertz probe spectroscopy (PPP-THz) and used it to uncover obscure aspects of hot carrier relaxation in the benchmark metal-halide perovskite (MHP) methylammonium lead-iodide. The technique allows us to systematically promote band-edge states in semiconductors by means of infrared optical excitation and monitor their subsequent relaxation with a non-contact probe of the photoconductivity. In the case of MAPbI$_3$, we observe a ~35% drop of the photoconductivity upon supplying the carriers with 600 meV of excess energy. The full recovery of this mobility occurs within a picosecond, but is postponed with respect to the population-only dynamics measured with an ultrafast infrared probe. We show that the observed dynamics can be modelled by a simple two-step sequence involving energy redistribution from the hot carrier to proximate strongly-coupled phonons, and subsequent thermal equilibration of the collective polaron-lattice state with the ambient environment. These results reveal the importance of local lattice heating on the mobility of energetic carriers in MHPs, outlining an additional framework for the design of hot carrier solar cells which consolidate the phononic and thermal properties of the absorber. We envisage that the PPP-THz technique will be instrumental in unlocking other niche optoelectronic applications of MHPs such as ultrafast optical switches,[76,77] as well as the characterization of fundamental excitations (excitons, carriers, charge-transfer states, polarons, trap states) in other emerging materials systems.

ASSOCIATED CONTENT

**Supporting Information**. The following supporting information is available free of charge:

Experimental details for sample preparation of $MAPbI_3$ films; experimental methods for steady-state and time-resolved spectroscopic measurements; linear optical properties of the $MAPbI_3$ films; instrument response function for the time-resolved THz setup; power spectrum of the THz probe pulses; fluence-dependent visible pump – THz probe data; visible pump – infrared push – THz probe kinetics for different pump-push delay times; effect of push fluence on the signal amplitude of the pump-push-probe experiments with NIR and THz probes; global fitting of the experimental data across a range of push fluences.

AUTHOR INFORMATION

**Author Contributions**

§ X.Z., T.R.H. and A.G. contributed equally to this work. T.R.H. and A.G. conceived the experiments. X.Z. and A.G. built the THz-PPP setup. T.R.H. and A.G. built the PPP-IR experiment. X.Z., T.R.H. and A.G. and performed the ultrafast measurements. X.Z., T.R.H., A.G., M.M. and A.A.B. analyzed the time-resolved data. B.A.A.M. and J.M.F. performed the conductivity calculations on the frequency-dependent data. W.X. prepared the samples. All authors commented on the final manuscript.

**Notes**

The authors declare no competing financial interests.

ACKNOWLEDGMENTS

We thank Prof. James Durrant for providing access to sample fabrication and handling facilities. This project received funding from the European Research Council (ERC) under the European Union's Horizon 2020 research and innovation programme (Grant Agreement No. 639750). T.R.H. acknowledges funding from an EPSRC Doctoral Prize Fellowship. A.A.B. and J.M.F. are Royal Society University Research Fellows (URF-R-191026 and URF-R1-191292 respectively). A.A.B. and X.Z. thank the Leverhulme Trust for support through a Philip Leverhulme Prize. This work utilized expertise and prototyping equipment at the Imperial College Advanced Hackspace.